# Dynamic and integrated mechanical movements of a rat brain associated with evoked potentials

Kunihiko Goto[a,*] and Toshio Nakaye[b]

[a]*Department of Molecular Pharmacology and Biological Chemistry, Northwestern University Medical School, 303 East Chicago Avenue, Chicago, IL 60611-3008, USA*
[b]*Department of Physiology, Tohoku University School of Medicine, 2-1, Seiryomachi, Aoba-ku, Sendai, 980 Japan.*

[*]*Corresponding Author*. Fax: +1-312-503-1700 E-mail address: k-goto@northwestern.edu (Kunihiko Goto)

**Abstract**

By using a piezoelectric sensor, it was demonstrated that the visual evoked potential of a rat brain was accompanied by mechanical movements of the brain when it was excited. A phase of upward movement was found to be followed by a phase of downward movement. The largest upward movement was a rise in swelling pressure on the order of 100 μg, which was about 40 times larger than that of the bullfrog sympathetic ganglion. The waves of mechanical movements were more complicated than those of the evoked potentials. These findings are thought to be due to the fact that the evoked potentials are propagated from the immediate surroundings of the sensor, while the mechanical signals are produced from anywhere beneath the piezo sensor. The mechanisms of mechanical movements propagated in the brain by electrical stimulation are discussed.

# Introduction

Evoked potentials are systemic events generated by the whole brain, and have been investigated to understand physiological pathways and states of the brain system. Within neural subsystems of the brain, various observations, which are considered to be directly associated with molecular conformational changes, have been reported in relation to molecular photochemistry [1]-[3], heat generation [4], [5], volume changes [6] and piezo sensor detection [7]. Tasaki et al. previously described the mechanical changes of nerve fibers and synapses, associated with action potentials [7]-]10]. Their findings that some components, i.e., nerve fibers and synapses, of the brain system move mechanically suggest that the brain system as a whole may exhibit mechanical movement. Thus, we tried to detect the mechanical movement of a rat brain while recording evoked potentials. For the demonstration of rapid mechanical changes associated with electric responses, substantially the same devices as those used in previous studies of the mechanical changes in the nerve fibers and synapses [7]-[12] were employed.

# Materials and Methods

**Perfusion of the rat brain**

Perfused rat brain was induced using a modification [13] of the procedure described by Andjus et al [14]. One day prior to the experiments, both vertebral arteries of a rat, weighing between 250 and 300 g, were electrocauterized with a monopolar coagulator at the first cervical vertebra. Under Nembutal (Sigma Chemical Co., St. Louis, MI, USA) and ether anesthesia, the perfusion fluid, perfluochemical blood substitutes, FC-43 Emulsion (Green Cross Corp., Osaka, Japan), mixed with a stream of gas (5% $CO_2$, 95% $O_2$), and monitored by a manometer (100 mmHg), were perfused into both pterygopalatine arteries, by way of both external carotid arteries. At the same time the external jugular veins were opened. At the time both respiration and heartbeat stopped. The rat was then fixed to a stereotaxic instrument.

**Detection of evoked potentials and mechanical changes**

The evoked potentials induced by the stimulation of the left optic nerve were recorded with a small silver ball, extracellularly [15]-[17]. Stimuli were 10 V in amplitude and 120 μsec in duration repeated 2.6 times per second. Mechanical changes in the brain were detected with a piezoelectric sensor (Gulton Industries, Inc.). (The same sensor was previously employed by Tasaki et al [7]-[12]. After opening the right occipital scalp, the stylus, fixed at the tip of the sensor, was brought into gentle contact with the cortex at the very near site, 1 mm forward, as the small silver ball, an evoked potential detector (Fig. 1).
All the measurements were carried out at 30~32 degrees Celsius.

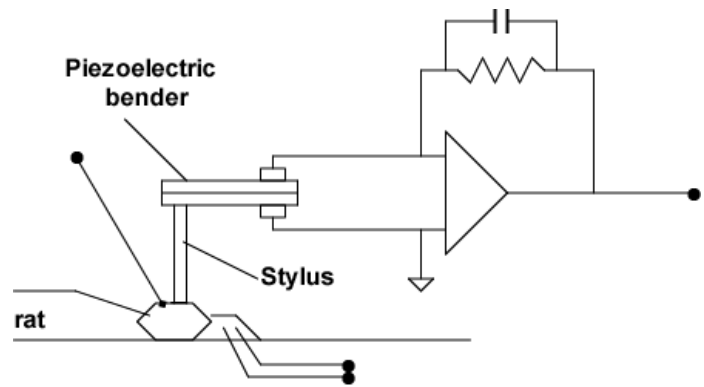

**Fig. 1**

# Results

Figure 2 shows an example of the records

(15 cases) obtained by the "piezoelectric" method of detecting rapid mechanical changes in a rat brain (lower traces) associated with the production of the evoked potential induced by the stimulation of the left optic nerve (upper traces). It is seen that a brief electric shock delivered to the nerve entering into the rat brain induced a first rapid upward deflection of the lower trace, followed by a second downward deflection. The upward deflection represents the appearance of an upwardly directed force on the order of 50 to 100 μg (81.9 $\pm$ 15.5 μg) (15 cases). This mechanical change started almost immediately or a fraction of a millisecond after the onset of the electrical response. In most cases, the mechanical signal consisted of several waves, of which the first wave was the largest. The wave images and the time course of the mechanical signal were similar to but more complicated (arrows of Fig. 2) than those of the evoked potentials. Similar phenomena in which the wave of the mechanical signals was more complicated than that of the evoked potentials was previously found in bullfrog sympathetic ganglion [9].

The findings that the mechanical signal was more complicated than the evoked signal can be explained by the fact that the electrical sensor detecting the evoked potential received only signals propagated near the tip of the evoked potential, while the piezo sensor detected all signals developed beneath its tip. The additional mechanical signals which are not found in the evoked signals may reflect the mechanical changes occurring in the lower part of the rat brain, beneath the area at which the evoked potential is detected. The difference between the mechanical changes and the evoked potentials may partly be due to the difference of detecting sites in the brain. Thus, the piezo sensor detector may be useful for detecting changes occurring at a deeper subcortical level in the brain.

The right traces of Fig. 2 show the mechanical movement and the evoked potential 45 minutes after perfusion. The first upward and second downward wave were still large, while the others became small. Though the duration of perfusion had been long, the mechanical changes were still more complicated than the evoked potential. Both the mechanical changes and the evoked potentials of the brain were observed for 2 hours after perfusion, while both the strength and the number of both mechanical movements and evoked potentials gradually became smaller.

At the peak of the mechanical signals, the mechanical increase determined by the piezoelectric sensor was between 50 and 100 μg. The size of the mechanical changes of bullfrog sympathetic ganglion was between 1 and 3 μg. The size of the mechanical changes in the rat brain was about 40 times that of the mechanical changes of bullfrog sympathetic ganglion detected by Tasaki et al [9].

The time required to fall to the resting level was usually between 300 and 400 ms.

## Discussion

We detected dynamic and integrated mechanical changes of the brain associated with the visual evoked potential. It is noteworthy that the brain employed was that of a mammal, the rat.

A variety of excitable cells were found to swell first and shrink later when they are excited. Tasaki et al. [7]-[10] have reported mechanical changes of nerve fibers and synapses, which are components of the brain system, and thus these components may cause mechanical changes directly to the rat brain or other components in addition to direct effects that may drive the mechanical changes, for example, nicotinic acetylcholine receptors, the swelling with large conformational changes which we previously described [18].

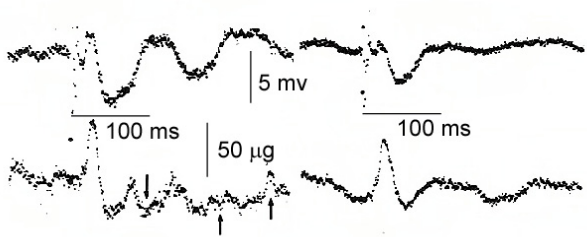

**Fig. 2**

The possible mechanism of the volume change is that inner hydrophobic sites of the acetylcholine receptor [18] and of underlining actin filaments [19] exposed in their functioning repel the surrounding water molecules, resulting in an increase in the volume corresponding to the repelling water molecules. A rat brain contains about $1 \times 10^8$ neurons, and a neuron has several thousand to several ten thousand spines (in this calculation, $3 \times 10^4$ was regarded as the number of spines per neuron [20]). One spine, that is synapse, contains a few thousand receptors of postsynaptic density (PSD) in the surface and underlining actin filaments. These functioning molecules, receptors [18] and actin filaments [19], expose their inner hydrophobic sites, removing surrounding water molecules, as described above. In order to calculate the swelling volume, we assume that the spine size is about 1 μm diameter and the outer side and inner side of surface receptors, and of actin filaments remove surrounding water molecules, each at 2 nm depth [7], [18]. Thus, the volume change per spine is $4 \times 0.002 \times 1$ (length; μm) $\times 1$ (transverse; μm) (μm$^3$). The swelling per neuron is $4 \times 0.002 \times 3 \times 10^4$ (μm$^3$). Thus, the swelling volume of one rat brain is $4 \times 0.002 \times 3 \times 10^4 \times 10^8$ (μm$^3$).

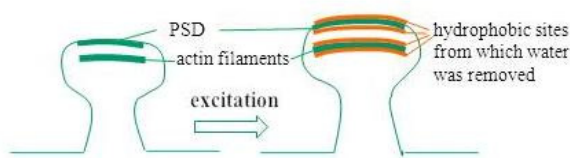

Fig. 3

As the volume of the rat brain is estimated as a 10 mm diameter globe, the volume change, C, is a solution to the following function:

$$4\pi/3 \times (5 + C)^3 = 4\pi/3 \times 5^3 + 4 \times 0.002 \times 3 \times 10^4 \times 10^8$$

$$C = 0.075 \text{ (μm)}$$

This value corresponds to the experimental volume change. Thus, the rat brain movement is suggested to occur by the exposure of hydrophobic sites of functioning macromolecules in the optic nerve excitation. Actually, the increase in size of spines immediately after excitation has been observed [20]. The mechanisms of size increase are unknown, but the mechanism described above would be reasonable.

Various cells and tissues move mechanically when they are excited. These phenomena are general and suggest that these mechanisms should be included in models of the interaction of macromolecules, e.g., in the "rotation model" [18], [19], [21]-[31] and corroborative evidence of this model recently [32].

Further experimental studies are required in order to clarify the significance of the phenomenon of swelling described in this report, which might be compared to swelling in fumarase catalysis [33], of polyelectrolyte gels [34], of artificial membranes [35], and of charged membranes [36], as studied by many investigators.

This study was carried out in accordance with the Guidelines for Animal Experimentation, Tohoku University and Tohoku University School of Medicine.

## Acknowledgements

This work was supported by a Grant-in-Aid of Scientific Research from the Ministry of Education, Science and Culture, Japan and by a Grant-in-Aid from Ono Pharmaceutical Co.,

Osaka. We thank Norman D. Cook for stimulating discussions.

**Figure legends**

Fig. 1. Schematic diagram illustrating the "piezoelectric" method of measuring mechanical changes in the rat brain. The stylus of the piezoelectric sensor pressing against the surface of the rat brain, the silver ball and stimulating electrodes are shown.

Fig. 2. The evoked potentials measured by the potential detector and the mechanical changes by the piezoelectric sensor. Signal-averager record shows the evoked potentials (upper trace) in a rat brain induced by electric stimulation of the optic nerve and mechanical changes (lower trace). Left: 15 minutes after perfusion. Right: 45 minutes after perfusion. Arrows are the additional mechanical changes suggested.

Fig. 3. The suggested mechanism of the volume change in a spine. It has been clarified that spines increase in size at excitation [37], however the mechanism of spine increase is unknown. We suggest that receptors in post synaptic density and underlining actin filaments of spines expose hydrophobic subsites of protein interior at excitation, and exposed hydrophobic subsites expel surrounding water molecules, making the volume increase.